\documentclass[12pt]{article}

\usepackage{graphicx}
\usepackage{amsmath}
\usepackage{amssymb}

\def\eq#1{(\ref{#1})}
\def\[#1\]{\begin{align}#1\end{align}}

\begin{document}

\begin{titlepage}
\title{
\hfill\parbox{4cm}{ \normalsize YITP-14-3 \\ WITS-CTP-128 }\\
\vspace{1cm} 
Ising model on random networks \\ and the canonical tensor model}
\author{
Naoki {\sc Sasakura}$^a$\thanks{\tt sasakura@yukawa.kyoto-u.ac.jp}
\and Yuki {\sc Sato}$^b$\thanks{\tt Yuki.Sato@wits.ac.za}
\\[15pt]
$^a${\it Yukawa Institute for Theoretical Physics, Kyoto University,}\\
{\it Kyoto 606-8502, Japan} \\
\\
$^b${\it National Institute for Theoretical Physics, }\\
{\it Department of Physics and Centre for Theoretical Physics,} \\
{\it University of the Witwartersrand, WITS 2050, South Africa}
}
%\date{\today}
%\normalsize}

\maketitle
\thispagestyle{empty}
\begin{abstract}
\normalsize
We introduce a statistical system on random networks of trivalent vertices 
for the purpose of studying the canonical tensor model, which is a rank-three tensor model in the canonical 
formalism. 
The partition function of the statistical system has a concise expression in terms of integrals, 
and has the same symmetries as the kinematical ones of the canonical tensor model.
We consider the simplest non-trivial case of the statistical system corresponding to the Ising model on random networks,
and find that its phase diagram agrees with what is 
implied by regrading the Hamiltonian vector field of the canonical tensor model with $N=2$
as a renormalization group flow.
Along the way, we obtain an explicit exact expression of the free energy of the Ising model on random networks
in the thermodynamic limit by the Laplace method.
This paper provides a new example connecting a model of quantum gravity and a random statistical system.
\end{abstract}
\end{titlepage}

\section{Introduction}
\label{sec:introduction}
The difficulty of combining
quantum mechanics and general relativity indicates our serious lack of 
a consistent description of nature.  In spite of this difficulty, quantum gravitational phenomena have been estimated 
through various thought experiments \cite{Garay:1994en}. 
They coherently suggest that the classical
notion of space-time as a smooth and continuous manifold should be 
replaced in some manner by a new quantum notion.
Then the mission to construct quantum gravity would be
to set up a dynamical theory of such a quantum notion and 
study its dynamics to see whether the common conventional characteristics of the space-time, such as 
smoothness, locality, classicality, causality, 
dimensionality, gravity, etc., can be derived as infrared emergent phenomena.  

The setup we consider in this paper is a rank-three tensor model \cite{ambjorn, sasakura, godfrey}
in the canonical formalism
\cite{Sasakura:2011sq,Sasakura:2012fb,Sasakura:2013gxg,Sasakura:2013wza},
dubbed {\it canonical tensor model} for short.
It is supposed to be a theory of dynamical fuzzy spaces \cite{Sasakura:2011ma}, 
and its dynamical variables are a canonical conjugate pair of tensors with three indices of
a certain cardinality\footnote{Namely, the indices can take $N$ values, say $1,2,\ldots,N$.}$N$.
The model is formulated as a totally constrained system with a number of first-class constraints
including Hamiltonian ones. Remarkably, the Hamiltonian constraints 
can be uniquely fixed by the algebraic consistency of the constraints and some other physically 
reasonable assumptions. 
Indeed, the theoretical structure of the model is very similar to the ADM formalism of general relativity
\cite{Arnowitt:1962hi,Dirac:1958sq,Dirac:1958sc,DeWitt:1967yk,Hojman:1976vp}.
On the other hand, our knowledge of the dynamics of the model is quite limited;
locality is favored as a result of quantum dynamics at least for $N=2$ 
\cite{Sasakura:2013wza}, and 
the mini-superspace approximation of general relativity in any dimensions 
can be derived from the model with $N=1$ \cite{Sasakura:2014gia}. 
Thus the present stage of its study is really far from deciding 
if the classical space-time can be derived as an infrared emergent phenomenon,
because of the lack of efficient methods to treat the model.

A possible strategy to study the dynamics of a model would be to relate it to a simpler system.
In this paper, a certain statistical system on random networks is considered as such a simpler system.
The random character of networks can be incorporated by defining the partition function of a system
in terms of a certain integral form. Then it is seen that the system has the same 
symmetries as the kinematical ones of the canonical tensor model. 
This may constrain the possible dynamics of the statistical system in some manner,
as the Hamiltonian constraint is determined uniquely in the canonical tensor model. 
In fact, we especially consider the case of the Ising model
on random networks, and have found 
that the phase structure is in remarkable agreement with
the prediction of the canonical tensor model with $N=2$, if the Hamiltonian vector field of the canonical
tensor model is regarded as the renormalization group flow of the Ising model.

We mention some former works about the Ising model on random graphs.
The Ising model on random trees is investigated in \cite{Durhuus:2011au}.
The Ising model on random surfaces is solvable in terms of 
the matrix model \cite{Kazakov:1986hu,Boulatov:1986sb}.
This is extended \cite{Bonzom:2011ev} to the colored tensor model \cite{Gurau:2009tw}, 
in which the generated graphs 
are supposed to be dominated by branched polymers
rather than higher-dimensional random manifolds \cite{Gurau:2011xp}. 
The random graphs we consider are different from the above works;
we consider networks generated by connecting trivalent vertices in absolutely random ways. 
The Ising model on Feynman diagrams, which are regarded as random networks as in the current work, 
is discussed in \cite{Bachas:1994qn,johnston} with a method different from ours.
The case with arbitrary degrees of vertices is treated 
in \cite{Dorogovtsev:2002ix,leone} by condensed-matter techniques.
There exist also some closely related works \cite{whittle,dembo} from mathematical viewpoints.

This paper is organized as follows.
In Section \ref{sec:isingspin}, we first present a grand type of partition function which
has a formal expansion in a dummy variable so that each order of the expansion 
gives the partition function of a statistical system
on random networks of the corresponding number of trivalent vertices.  
The symmetries of the system are discussed.
In Section \ref{sec:perturbation}, as the simplest non-trivial case of the statistical model,
we study the phase structure of the Ising model 
on random networks by numerically evaluating the polynomial function expressing the partition function. 
We see that the phase diagram agrees with the prediction made by regarding 
the Hamiltonian vector field of the canonical tensor model with $N=2$ 
as the renormalization group flow of the Ising model.
In Section \ref{sec:meanfield}, we apply the Laplace method to obtain an explicit exact expression 
of the free energy of the Ising model on random networks in the thermodynamic limit. 
The result agrees with the study in Section \ref{sec:perturbation}.
In Section \ref{sec:nonperturbative}, 
we discuss the non-perturbative definition of the grand partition function,
and, in Section \ref{sec:numnon}, study its singular property in the non-perturbative regime.
We find an agreement with a prediction based on the orientation-reversed renormalization group flow.   
In Section \ref{sec:tensormodel}, we derive the Hamiltonian vector field
of the canonical tensor model. 
The final section is devoted to a summary and consideration of future prospects.

\section{A statistical system on random networks of trivalent vertices}
\label{sec:isingspin}
The statistical model considered in this paper is generated from the following grand type of partition function,
\[
Z(M,t)=\int \prod_{d=1}^N d\phi_d\ e^{- \phi_a \phi_a+\, t \, M_{abc} \phi_a \phi_b \phi_c},
\label{eq:partdef}
\]
where the indices take integer values from 1 to $N$, and we take the convention that 
repeated indices are supposed to be summed over, unless otherwise stated.
Here $M_{abc}$ is a real symmetric tensor with three indices, 
and $t$ is a dummy variable introduced for convenience.
To define the grand 
partition function \eq{eq:partdef} properly, a special care of the integration contours of $\phi_a$
is needed due to the unboundedness of the triple coupling term. 
This will be discussed in Section \ref{sec:nonperturbative}.

Here let us first treat the grand partition function $Z(M,t)$ in perturbation of $t$. 
A formal expansion is given by
\[
Z(M,t)\simeq \sum_{n=0}^\infty t^n Z_n(M)=\sum_{n=0}^\infty \frac{t^n}{n!} 
\int_{-\infty}^\infty \prod_{d=1}^N d\phi_d\ (M_{abc} \phi_a \phi_b \phi_c)^n 
e^{-\phi_a\phi_a}.
\label{eq:pert}
\]
In this perturbative treatment, the integration contours are set over the real axes,\footnote{This will be justified in 
Section \ref{sec:nonperturbative}.}
and the integrations are well defined for each $Z_n(M)$.
By employing the Wick theorem for the Gaussian integrations, 
$Z_n(M)$ can be described graphically by the summation of the Feynman diagrams
which are closed networks randomly connecting $n$ trivalent vertices.  
Note that $Z_{n={\rm odd}}=0$ for trivalent vertices.  
Note also that such networks contain disconnected ones in general, 
but actually such contributions are negligible in the thermodynamic limit $n\rightarrow \infty$, which will be discussed later. 
This can be checked explicitly by computing  the $N=1$ case, 
which has only the graphical degrees of freedom, and comparing the perturbative expansions in $t$ of 
$\frac{1}{\sqrt{\pi}}Z(M,t)$ and 
$\log \frac{1}{\sqrt{\pi}}Z(M,t)$, where the coefficients of the latter expansion count only the connected diagrams 
as textbook knowledge.
Thus the networks representing $Z_n(M)$ can effectively be regarded as connected ones 
with probability 1 in the large $n$ limit.

In addition to the graphical aspect above, 
each vertex of a Feynman diagram is weighted by $M_{abc}$, and 
the summation over the indices can be regarded as that over the degrees of 
freedom on edges. 
Thus $Z_n(M)$ can be regarded as the partition function of a statistical system on 
random networks
which have $n$ trivalent vertices and one degree of freedom, which can take $N$ different values, on each edge.

In this paper, we will consider explicitly the case of $N=2$, which corresponds to the Ising model on random networks. 
The relation between $M$ and the common thermodynamic variables is given by
\[
M_{abc}=e^{\frac{H}{2}(\sigma_a +\sigma_b+\sigma_c)+
J (\sigma_a \sigma_b+\sigma_b \sigma_c + \sigma_c \sigma_a)
+K \sigma_a \sigma_b \sigma_c},
\]
where $\sigma_a$ takes the values of spin up and down,
$\sigma_1=1,\ \sigma_2=-1$, and 
$H$, $J$ and $K$ represent a magnetic field, a nearest-neighbor coupling, and
a triple-coupling, respectively. 

Alternatively, one can put Ising spins on vertices.
In this case, the relation to the thermodynamic variables is given by
\[
M_{abc}=\sum_{i=1}^2 R_{ai}R_{bi}R_{ci} e^{H \sigma_i}, 
\label{eq:relMR}
\]
where $R$ is a two-by-two real matrix satisfying 
\[
(R^T R)_{ij}=e^{J \sigma_i \sigma_j}.
\label{eq:relRHJ}
\]
For the ferromagnetic case $J\geq 0$, it is possible to find such a real $R$.

The partition function $Z_n(M)$ defined in \eq{eq:pert} is invariant under the 
orthogonal group transformation, $L\in O(N)$, as
\[
Z_n(L(M))=Z_n(M),
\label{eq:invorthopert}
\]
where 
\[
L(M)_{abc}=L_{aa'}L_{bb'}L_{cc'} M_{a'b'c'}.
\]
This invariance can be proved by using the invariance of the integration measure in \eq{eq:pert}
under the orthogonal group transformation of $\phi$, $\phi_a'=\phi_b L_{ba}$.
Similarly, for this symmetry to be satisfied also by the grand partition function \eq{eq:partdef},
one has to define the integration measure in \eq{eq:partdef} properly. 
In fact, there exists a simple choice of the integration contours satisfying this invariance, as will be discussed in 
Section \ref{sec:nonperturbative}, and we have 
\[
Z(L(M),t)=Z(M,t).
\label{eq:invortho}
\]

The grand partition function \eq{eq:partdef} also has a rescaling invariance given by
\[
Z(e^{\sigma} M,e^{-\sigma} t)=Z(M,t)
\label{eq:invscale}
\] 
with real $\sigma$.
By ignoring the transformation of the dummy variable $t$, 
these orthogonal and scale transformations of $M$ can be considered
to define the gauge transformations of the statistical model.
In fact, there is a physical reason to regard the scale transformation as an invariance:
under the scale transformation, each $Z_n(M)$ is transformed by an overall factor, which is 
irrelevant for the statistical properties.

In the present case of $N=2$, these gauge transformations delete two of the degrees of freedom of $M$,
which originally has four independent tensor elements.
A convenient choice of gauge is given by \cite{Sasakura:2013wza}
\[
M_{111}=1,\ M_{112}=0,\ M_{122}=x_1,\ M_{222}=x_2,
\label{eq:gauge}
\]
where the $x_i$s are real. Though this gauge is convenient for explicit computations,
it suffers from the so-called Gribov ambiguity, namely, in general there exist distinct points in the $(x_1,x_2)$ space,
which are actually gauge equivalent. To distinguish the gauge equivalence/inequivalence, 
it is convenient to consider for instance the following gauge-invariant quantities,
\[
Q_1(M)&=\frac{g_1(M)}{g_0(M)}, \label{eq:Q1}\\
Q_2(M)&=\frac{g_2(M)}{g_0(M)^2}, \label{eq:Q2}
\]
where the $g_i$s are the $O(N)$-invariant quantities defined by 
\[
\label{eq:g0}
g_0(M)&=M_{abc}M_{abc},\\
\label{eq:g1}
g_1(M)&=M_{aab}M_{bcc},\\
\label{eq:g2}
g_2(M)&=M_{abc}M_{bcd}M_{def}M_{efa}.
\]
By evaluating the values of $Q_i$, one can tell the gauge equivalence/inequivalence of 
the points in the $(x_1,x_2)$ space.\footnote{One may consider additional gauge-invariant quantities 
to reject the possibility that $Q_i$'s accidentally coincide for gauge-inequivalent points.}

\section{Numerical study of the partition function of the Ising model on random networks }
\label{sec:perturbation}
Let us set $N=2$ and consider the gauge \eq{eq:gauge}. The partition function $Z_{2n}$ in \eq{eq:pert} is given by
\[
Z_{2n}(x_1,x_2) &= \frac{1}{(2n)!} \int_{-\infty}^\infty d{r_1} d{r_2} (r_1^3+3 x_1 r_1 r_2^2+x_2 r_2^3)^{2n} 
e^{-r_1^2-r_2^2} \\
&= \sum_{l_1,l_2=0}^{l_1+2 l_2\leq 2 n}
\frac{\left(3n- l_1-3 l_2-\frac{1}{2}\right)!\left(l_1+3 l_2 -\frac{1}{2}\right)! }{(2n-l_1-2l_2)! ( l_1)! (2 l_2)!} (3 x_1)^{ l_1} x_2^{2l_2}.
\label{eq:part2nx1x2}
\]
Here, from the first to the second line, we have expanded 
$(\cdots)^{2n}$ in polynomials and carried out the Gaussian integrations of all the terms.
Let us consider the free energy per vertex defined by
\[
f_{2n}(x_1,x_2)=-\frac{1}{2n} \log\left[Z_{2n}(x_1,x_2)\right].
\]

\begin{figure}[]
\begin{center}
\includegraphics[]{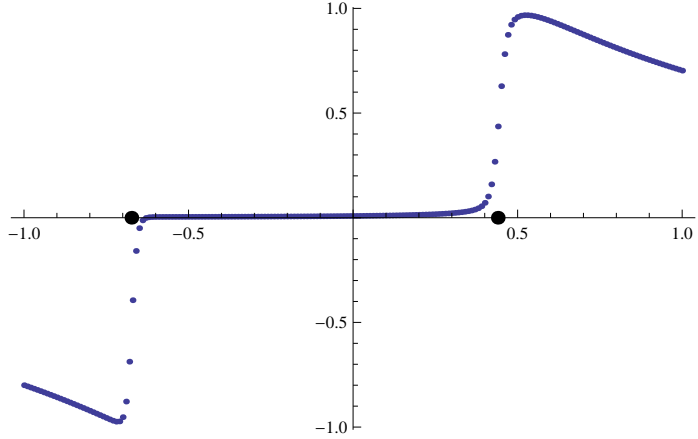}
\caption{Plot of $-\partial_{x_1} f_{100}(x_1,x_2)$ for
$-1\leq x_1 \leq 1$ with intervals $0.01$, and $x_2=0.5$.
The horizontal axis represents $x_1$.
The larger dots on the horizontal axis 
show the locations of the phase transition lines \eq{eq:transpoints} for $x_2=0.5$ 
predicted from the tensor model.}
\label{fig:x205}
\end{center}
\end{figure}

\begin{figure}[]
\begin{center}
\includegraphics[]{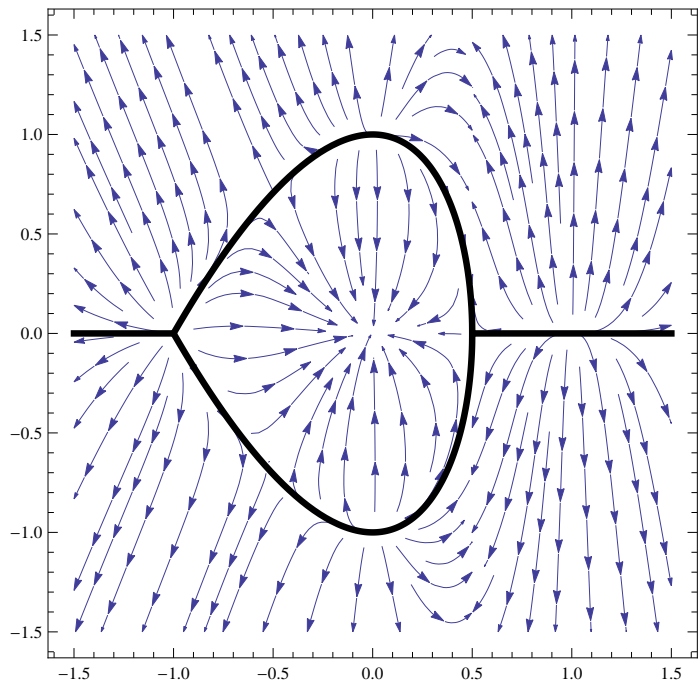}
\caption{
The Hamiltonian vector field of the canonical tensor model in the $(x_1,x_2)$ space.
The horizontal and vertical axes represent $x_1$ and $x_2$, respectively.
The solid lines are the expected phase transition lines.
\label{fig:rgflow}}
\end{center}
\end{figure}

Since \eq{eq:part2nx1x2} is merely a polynomial function of $x_i$s, one can easily study numerically 
the properties of $f_{2n}(x_1,x_2)$ in various regions of the parameters.
For instance, Figure \ref{fig:x205} plots $-\partial_{x_1} f_{100}(x_1,x_2)$ for
$-1\leq x_1 \leq 1$ and $x_2=0.5$.
One can clearly see that there exist jumps around $x_1\sim -0.7$ and $x_1\sim 0.4$.
In fact, the jumps become sharper for larger $n$, while
their locations and the values of $-\partial_{x_1} f_{2n}(x_1,x_2)$
around them seem to converge for larger $n$.
These facts indicate the existence of first-order phase transitions at these points 
in the thermodynamic limit $n\rightarrow \infty$.
By studying various parameter regions, one can find that the locations of the phase transition lines
are consistent with the solid lines in Figure \ref{fig:rgflow}.
The solid lines are the expected phase transition lines 
derived by regarding the Hamiltonian vector field of the canonical tensor model,
which will be discussed in Section \ref{sec:tensormodel}, as a renormalization group flow. 
The equations of the lines are given by
\[
x_2=\left\{
\begin{array}{cl}
\pm (1 + x_1)\sqrt{1 - 2 x_1},& -1\leq x_1 \leq \frac{1}{2},\\
0,& \hbox{otherwise}. \\
\end{array}
\right.
\label{eq:transpoints}
\]
In fact, the values of $x_1$ for $x_2=0.5$ are indicated by the larger points 
on the horizontal axis in Figure \ref{fig:x205}, and 
their locations are in good agreement with the jumping points.

\begin{figure}[]
\begin{minipage}{8cm}
\includegraphics[width=8cm]{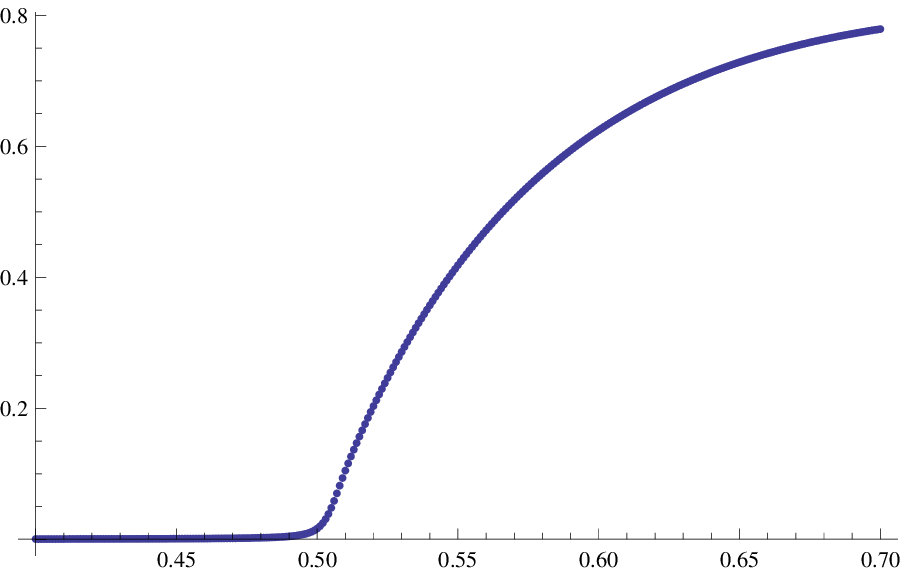}
\end{minipage}\hspace{2pc}%
\begin{minipage}{8cm}
\includegraphics[width=8cm]{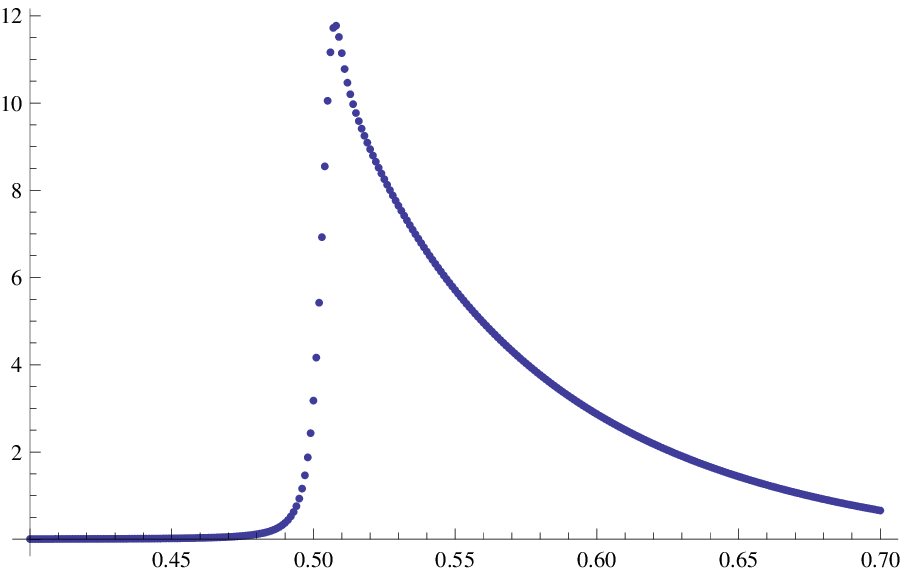}
\end{minipage} 
\caption{The left and right figures show $-\partial_{x_1} f_{10000}$ and $-\partial_{x_1}^2 f_{10000}$,
respectively,  for $0.4\leq x_1 \leq 0.7$ with interval $0.001$, and $x_2=0$.
The horizontal axes represent $x_1$.}
\label{fig:qurie}
\end{figure}

The phase transitions seem to be mostly first order, 
but there also exists a second-order phase transition point at $(x_1,x_2)=(0.5,0)$. This can be
seen in Figure \ref{fig:qurie}, 
which plots $-\partial_{x_1} f_{10000}$ and $-\partial_{x_1}^2 f_{10000}$ for 
the interval $0.4 \leq x_1 \leq 0.7$ and $x_2=0$.  
While the first derivative is continuous, the second derivative shows a discontinuity
at $x_1=0.5$. 

The location of the second-order phase transition point $(x_1,x_2)=(0.5,0)$ agrees with the known result in
the literature \cite{Bachas:1994qn,johnston,Dorogovtsev:2002ix, leone}.
Generally for arbitrary degrees of vertices, the Curie point of the Ising model on random networks
has been argued to be given by  
\[
J_c=\frac{1}{2} \log\left[ \frac{\langle k^2 \rangle}{\langle k^2 \rangle -2 \langle k \rangle}\right],
\label{eq:critical}
\]
where $k$ and $\langle \ \rangle$ denote degrees of vertices and average over networks, respectively. 
Since, in our case, all the vertices are trivalent, the critical coupling should be given by
\[
J_c=\frac{1}{2} \log[3].
\label{eq:criticalval}
\]

As discussed in Section \ref{sec:isingspin}, we can take two distinct interpretations concerning the positions 
of the degrees of freedom on networks. The above known result assumes that the Ising spins are 
on the vertices, and we also take this interpretation. Then, for $H=0$, 
the gauge-invariant quantities \eq{eq:Q1} and \eq{eq:Q2} take
\[
Q_1&=\frac{e^J + e^{3 J}}{e^{-3J} + e^{3 J}}=\frac{(x_1+1)^2+x_2^2}{3 x_1^2+x_2^2+1},\\
Q_2&=\frac{4 +  e^{-6 J} +  e^{-2 J} +  e^{2 J} + 
  e^{6 J} }{2(e^{-3J} + e^{3 J})^2}
  =\frac{2 x_1^2 x_2^2+\left(2x_1^2+x_2^2\right)^2+\left(x_1^2+1\right)^2}{\left(3 x_1^2+x_2^2+
1\right)^2},
\]
for \eq{eq:relMR}, \eq{eq:relRHJ} and for the gauge \eq{eq:gauge}.
These equations can be solved by
\[
(x_1,x_2)=(\tanh[J],0).
\]
Therefore \eq{eq:criticalval} is gauge-equivalent to $(x_1,x_2)=(\frac{1}{2},0)$ as expected.

\section{The exact free energy in the thermodynamic limit}
\label{sec:meanfield}
In fact, the free energy $f_{2n}=-\frac{1}{2n} \log[Z_{2n}]$ can be 
computed exactly in the thermodynamic limit $n\rightarrow \infty$
in the following manner.
Let us first rewrite the partition function defined in \eq{eq:pert} as
\[
Z_{2n}(M)&= \frac{1}{(2n)!}\int_{-\infty}^\infty \prod_{d=1}^N d\phi_d\ (M_{abc} \phi_a \phi_b \phi_c)^{2n} 
e^{-\phi_a\phi_a} \nonumber \\
&=\frac{1}{(2n)!}\int_{-\infty}^\infty \prod_{d=1}^N d\phi_d\ 
e^{-\phi_a\phi_a+n \log [(M_{abc} \phi_a \phi_b \phi_c)^2]} \nonumber \\
&=\frac{(2n)^\frac{N}{2}}{(2n)!} \int_{-\infty}^\infty \prod_{d=1}^N d\phi_d\ 
e^{2n(-\phi_a\phi_a+\frac{1}{2} \log [(M_{abc} \phi_a \phi_b \phi_c)^2])+3n \log[2n] },
\label{eq:partintres}
\]
where we have rescaled $\phi \rightarrow \sqrt{2n}\phi$ in the last line.
Then, we can see that, in the thermodynamic limit $n\rightarrow \infty$, the free energy 
 is given by
\[
f(M)= \bar \phi_a \bar \phi_a- \frac{1}{2} \log[(M_{abc} \bar \phi_a \bar \phi_b \bar \phi_c)^2]+f_{nw},
\label{eq:freemean}
\]
where $f_{nw}=-\frac{1}{2} \log[2 n] -1$ is a divergent but $M$-independent part, 
and $\bar \phi$ is real and taken so that 
it gives the minimum of the right-hand side of \eq{eq:freemean} for a given 
$M$.\footnote{In the evaluation of the partition function in \eq{eq:partintres}, it is obvious that there exists also a 
contribution from $\phi=-\bar\phi$ as equally as $\phi=\bar \phi$. 
This results in a multiplicative factor of 2 to the partition function.
But this factor is irrelevant in the thermodynamic limit,
since \eq{eq:freemean} is affected merely by $O(\frac{1}{n})$.}
This is an application of the Laplace method to evaluate a real integration.
$f_{nw}$ would be regarded as the free energy of the networks rather than that of the statistical 
system on them.
The expression \eq{eq:freemean} has at most $\log[n]/n$ corrections, 
and is therefore exact in the thermodynamic limit.
Thus the system is treatable by a sort of mean-field-like method, and this would be consistent with the former
arguments on the critical exponents \cite{Dorogovtsev:2002ix,leone}.

The maximization condition on $\bar \phi_a$ may be solved by one of the solutions to the extremeness condition,
\[
2 \bar \phi_a-\frac{3 M_{abc} \bar \phi_b \bar \phi_c}{M_{abc} \bar  \phi_a \bar \phi_b \bar \phi_c}=0.
\label{eq:extremal}
\]
By rescaling $\bar \phi_a=g w_a$ with $g=M_{abc} \bar \phi_a \bar \phi_b \bar  \phi_c$,
\eq{eq:extremal} is equivalently given by
\[
2 w_a- 3 M_{abc} w_b w_c&=0,
\label{eq:Mw}\\
g^2 w_a w_a &= \frac{3}{2}.
\label{eq:Mg}
\]
With this parametrization, the free energy \eq{eq:freemean} is given by
\[
f(M)=\frac{3}{2} -\frac{1}{2}\log\left[ \frac{3}{2} \right]+\frac{1}{2} \log[w_a w_a ]+f_{nw}.
\label{eq:fgexp}
\]
Here we have used 
\[
\phi_a\phi_a=g^2 w_a w_a=g^2 w_a \cdot \frac{3}{2} M_{abc} w_b w_c=\frac{3}{2},
\]
which can be derived from \eq{eq:Mw} and $g=M_{abc} \bar \phi_a \bar \phi_b \bar\phi_c=g^3 M_{abc}w_a w_b w_c$.
Thus the problem of obtaining the exact free energy reduces to solving \eq{eq:Mw} and taking, 
from all the solutions, the one which minimizes $w_a w_a\ (>0)$. 

Now let us explicitly consider the Ising model $N=2$, and take the gauge \eq{eq:gauge}. 
Then the explicit form of \eq{eq:Mw} is given by
\[
&-2 w_1 + 3 w_1^2 + 3 x_1 w_2^2 =0, \\
&w_2 (-2 + 6 x_1 w_1  + 3 x_2 w_2 )=0.
 \]
There are two sets of meaningful solutions $(w_a w_a>0)$ as
\[
w_1=\frac{2}{3},\ w_2=0,
\]
and 
\[
w_1&= \frac{4 x_1^2 +  x_2^2 \pm  x_2 \sqrt{-4 x_1 + 8 x_1^2 + x_2^2}}{3(4 x_1^3 + x_2^2)},\\
w_2&=\frac{2 \left(x_2 - x_1 x_2  \mp x_1 \sqrt{-4 x_1 + 8 x_1^2 + x_2^2}\right)}{3 (4 x_1^3 + x_2^2)}.
\] 
By considering which of the solutions minimizes $w_a w_a$ for each $(x_1,x_2)$,
it is concluded that we should take
\[
w_a w_a=\left\{
\begin{array}{l}
{\displaystyle \frac{4}{9} }\hbox{ for  }
-1\leq x_1 \leq \frac{1}{2}, \ |x_2| <(1 + x_1)\sqrt{1 - 2 x_1}, \\
{\displaystyle \frac{
8 (1 - 6 x_1 + 9 x_1^2 + x_2^2)}{9 ( |x_2| (-4 x_1 + 8 x_1^2 + x_2^2)^\frac{3}{2}
-8 x_1^3 + 24 x_1^4 + 2 x_2^2 - 6 x_1 x_2^2 + 12 x_1^2 x_2^2 + x_2^4)}}\ 
\hbox{otherwise}.
\end{array}
\right.
\] 
This gives the exact free energy through \eq{eq:fgexp}, and 
confirms the phase structure and the numerical values in Section \ref{sec:perturbation}.

The gauge \eq{eq:gauge} is convenient for explicit computations and comparison with the canonical tensor model, 
but it turns out that it is not so for the study of 
the thermodynamical properties of the Ising model with the usual thermodynamic variables.
We will study the exact thermodynamic properties of the Ising model on random networks 
in a separate publication \cite{Sasakura:2014yoa}. 

\section{Non-perturbative definition of the grand partition function}
\label{sec:nonperturbative}
The grand partition function \eq{eq:partdef} is not well defined if the integrations of the $\phi_a$s are taken to be 
from $-\infty$ to $\infty$ on the real axis, because of the unboundedness of the triple coupling.
On the other hand, we should keep the perturbative properties discussed in Section \ref{sec:perturbation},
where the integrations are assumed so.
In fact, there exists a proper way to take the integration contours in complex planes to 
reconcile the two requirements.

For illustration, let us first discuss the simplest case with $N=1$. The grand partition function \eq{eq:partdef} 
is given by
\[
Z^{N=1}(M,t)=\int d\phi\ e^{-\phi^2+t M \phi^3}. 
\]
Let us assume $t,M>0$.
Then, to make the integration finite, one has to deform the contour from the real axis in such a way
that $\hbox{Re}(-\phi^2+t M \phi^3)\rightarrow -\infty$ for $|\phi|\rightarrow \infty$. 
A simple deformation would be
\[
\phi=e^{i\theta} r 
\hbox{  with }
\left\{
\begin{array}{cl}
-\frac{\pi}{6}\leq \theta \leq \frac{\pi}{6} & r \leq 0, \\
\frac{\pi}{6}\leq \theta < \frac{\pi}{2}& r >0.
\end{array}
\right.
\label{eq:deformN=1}
\]
Because of the fast damping of the integrand at infinite $|r|$, the partition function is independent of 
$\theta$ in the range.  

On the other hand, the perturbative expansion of the grand partition function in $t$ is given by 
\[
Z^{N=1}(M,t)\simeq\sum_{n=0}^\infty \frac{(tM)^n}{n!} \int d\phi\ \phi^{3n} e^{-\phi^2}.
\label{eq:partN=1}
\]
In Section \ref{sec:perturbation}, the integrations are assumed to be over the real axis from 
$-\infty$ to $\infty$.
In fact, because of the fast damping of the integrands, the Gaussian integrals of each order term in
\eq{eq:partN=1} do not 
depend on the change of contours in the form $\phi=e^{i\theta} r$ with $-\frac{\pi}{4}<\theta<\frac{\pi}{4}$. 
Since this range contains the real axis and also overlaps with the region in \eq{eq:deformN=1}, 
it is concluded that the perturbative property of the grand partition function with 
the integration contour \eq{eq:deformN=1} is consistent with the treatment in 
Section \ref{sec:perturbation}.

Once the grand partition function is defined for $t,M>0$, one can define it for other parameter 
regions by analytic continuation. 
This is equivalent to deforming the integration contour continuously so that 
$\hbox{Re}(-\phi^2+t M\phi^3)\rightarrow -\infty$ for $|\phi|\rightarrow \infty$
is forced to hold throughout the continuous change of $t,M$ in complex values.
With this procedure, one can easily show that the grand partition function can be equivalently defined 
by considering first pure imaginary $t M$ with the integration of $\phi$ over the real axis from $-\infty$ to $\infty$
and then performing the analytical continuation of $tM$ to real 
values.\footnote{All the above discussions for $N=1$ can be supported by the explicit expression of the grand partition function,
$Z^{N=1}(M,t)=\frac{2 i}{3 \sqrt{3}t M}e^{-\frac{2}{27 (tM)^2}}K_{\frac{1}{3}}\left(-\frac{2}{27 (tM)^2}\right)$,
with a modified Bessel function.}

The simplest choice in \eq{eq:deformN=1} would be $\theta=\frac{\pi}{6}$ commonly for both positive and negative $r$. 
In this case, the triple coupling takes pure imaginary values, and
the integration in \eq{eq:partN=1} is obviously convergent.
This is true for any $N$, and the contours can be taken to be
\[
\phi_a=e^{\frac{\pi}{6}i} r_a,\ (a=1,2,\ldots,N), 
\label{eq:nonpertcontour}
\]
with real $r_a$.
This simplest common choice of the contours guarantees the gauge invariance \eq{eq:invortho} of the grand 
partition function. 
More complicated choices of contours depending on $\phi_a$ or on the signature of $r_a$ 
will make the invariance a complicated issue to be proved, since the orthogonal transformation
\eq{eq:invortho} mixes the $\phi_a$s and also the positive and negative values of the $r_a$s.

Since the integration contours of $\phi_a$ are deformed from the real axes into the complex planes, the grand partition
function takes complex values in general even for real $t,M$. Since the perturbative part is the same as the
one obtained by the Gaussian integrations over the real axes, imaginary values appear 
only in the non-perturbative corrections to the grand partition function. We leave the physical interpretation of 
this aspect for future study.

\section{Numerical analysis of the grand partition function} 
\label{sec:numnon}
The analysis of the partition function $Z_{2n}(M)$ in Section \ref{sec:perturbation} concerns the perturbative 
aspect of the grand partition function $Z(M,t)$ about $t=0$. A new aspect of the non-perturbative definition
in the previous section would appear in its large $t$ limit.    
In this section, we will analyze this limit in the case of $N=2$ with the gauge \eq{eq:gauge}.

\begin{figure}[]
\begin{center}
\includegraphics[]{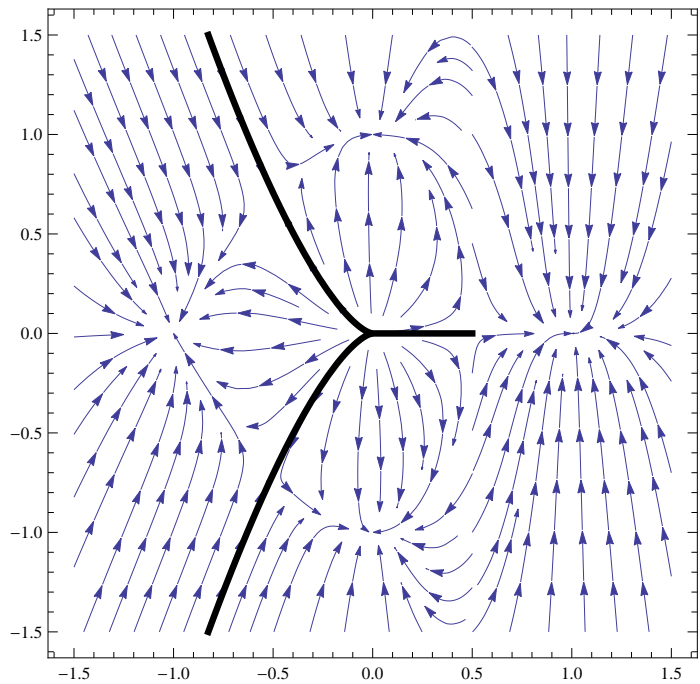}
\caption{The reversed Hamiltonian vector field of the canonical tensor model. The solid lines are 
the expected phase transition lines. 
The line $(x_1,x_2)=(\frac{1}{2},\hbox{arbitrary})$ is a locus of gauge singularities and is not physical.
}\label{fig:RGrev}
\end{center}
\end{figure}

The phase structure in such a limit could be predicted from the canonical tensor model
by reversing the Hamiltonian vector field 
employed in Figure \ref{fig:rgflow}. 
The reversed flow diagram is shown in Figure \ref{fig:RGrev}, and the solid lines are the expected phase 
transition lines,
which are given by
\[
\begin{array}{cl}
4 x_1^3+x_2^2=0, & x_1\leq 0, \\
x_2=0, & 0<x_1\leq 1/2.
\end{array}
\label{eq:nonperttrans}
\]

\begin{figure}[]
\begin{minipage}{8cm}
\includegraphics[width=8cm]{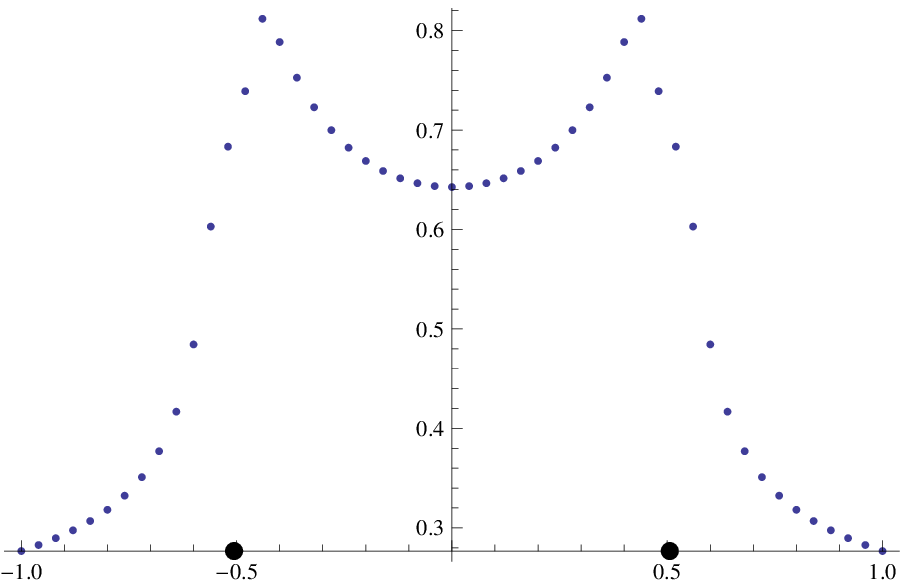}
\end{minipage}\hspace{2pc}%
\begin{minipage}{8cm}
\includegraphics[width=8cm]{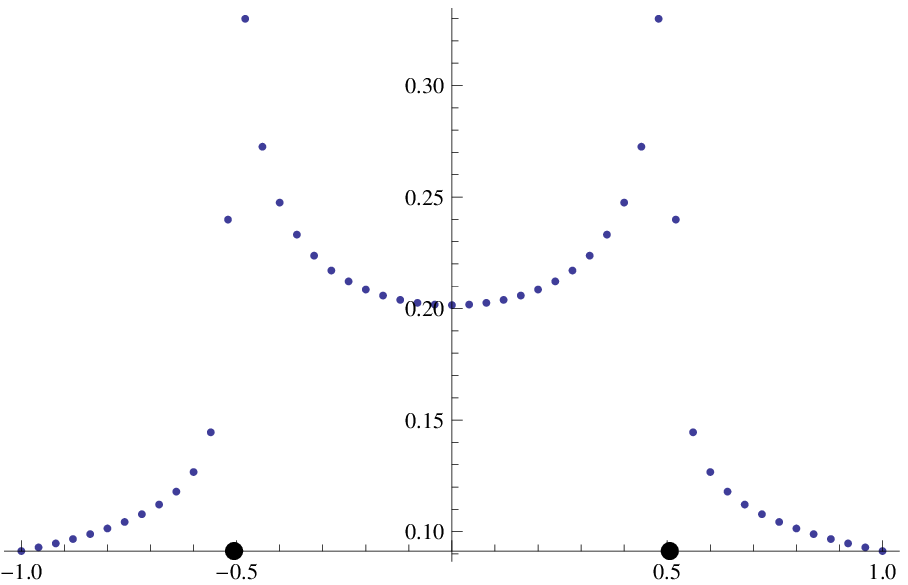}
\end{minipage} 
\caption{The real part of the grand partition function $Z(x_1,x_2,t)$ 
for $x_1=-0.4$ and $-1\leq x_2 \leq 1$ 
with interval $0.04$. The horizontal axes represent $x_2$. 
The left and right figures are for $t=10$ and $t=50$, respectively.
The larger points on the horizontal axes indicate the values of $x_2$ of the transition lines \eq{eq:nonperttrans}
at $x_1=-0.4$. }
\label{fig:x104}
\end{figure}

We will show below the results of the numerical analysis of the grand partition function.
Figure \ref{fig:x104} plots the 
real part of the grand partition function for $x_1=-0.4$ and $-1\leq x_2 \leq 1$. 
The left and right figures are for $t=10$ and $t=50$, respectively. 
The larger points on the horizontal axes indicate the 
values of $x_2$ of the phase transition lines \eq{eq:nonperttrans} at $x_1=-0.4$. 
At least in this range of values of $t$, the peaks are getting more enhanced by taking $t$ larger,  
and the peaks can be expected to become singularities as $t\rightarrow \infty$.
Their locations are consistent with \eq{eq:nonperttrans}. We have also studied some cases with different $x_1<0$,
and have obtained results consistent with \eq{eq:nonperttrans}. 

\begin{figure}[]
\includegraphics[width=5cm]{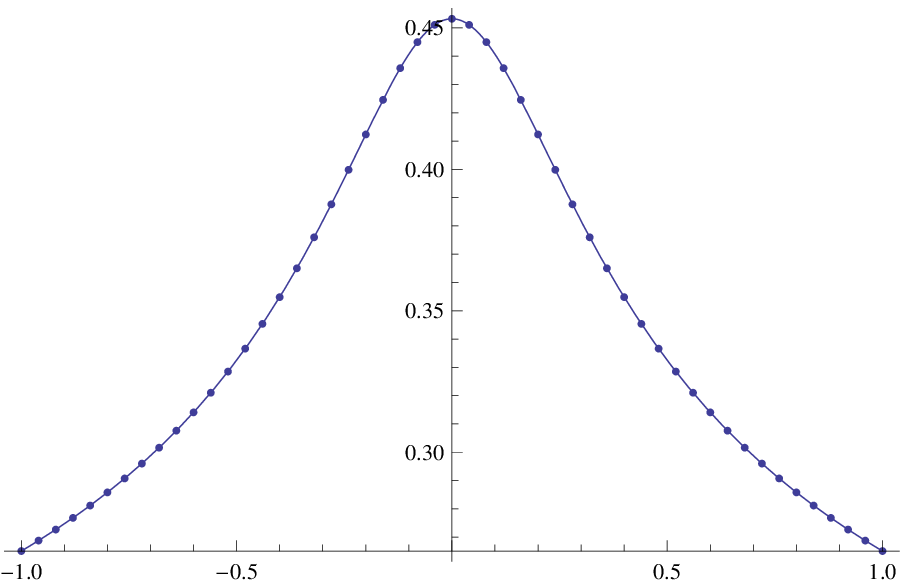}
\hfil
\includegraphics[width=5cm]{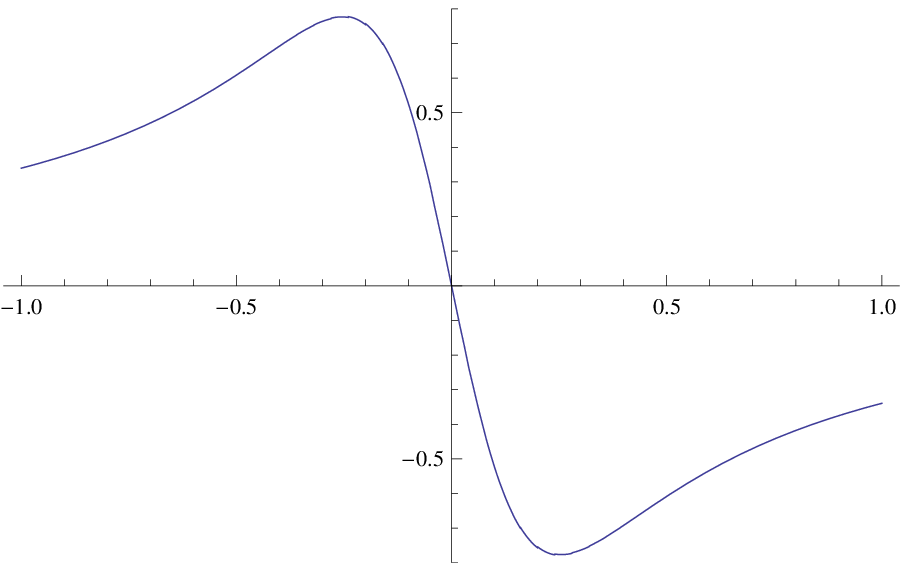}
\hfil
\includegraphics[width=5cm]{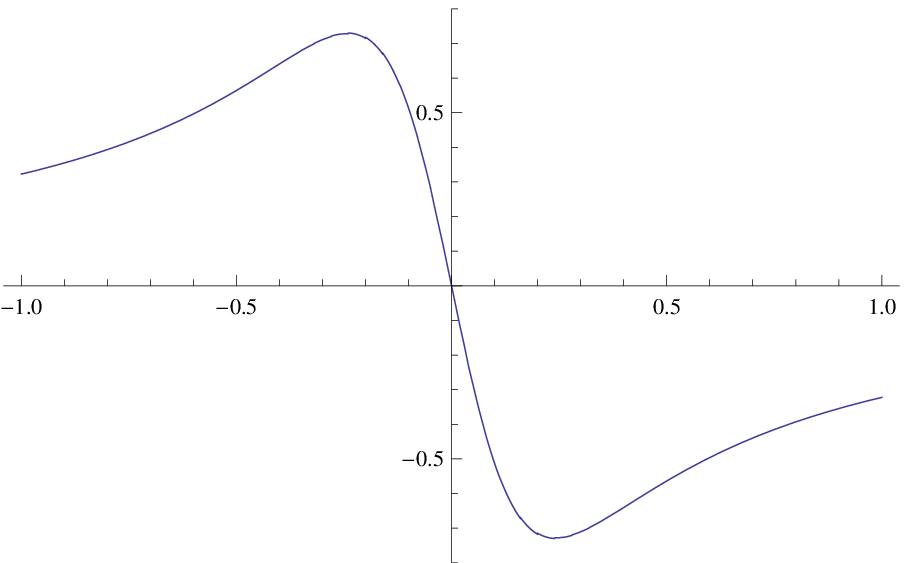}
\caption{In the left figure, the real part of $Z(x_1,x_2,t)$ is plotted for $t=10$, $x_1=0.25$ and $-1\leq x_2 \leq 1$ with interval $0.04$,
and the solid line represents an interpolating function $g(x_2)$.
The middle and right figures show $\frac{d }{dx_2} \log[g(x_2)]$ for $t=10$ and $t=50$, respectively. 
The horizontal axes represent $x_2$.}
\label{fig:t10x1025}
\end{figure}

Figure \ref{fig:t10x1025} compares the real part of the grand partition function for 
two values of $t$, $t=10$ and $t=50$, at $x_1=0.25$ and $-1\leq x_2 \leq 1$.
The values are interpolated by functions, and the first derivatives of their logarithms are shown
for each case of $t$. 
Though there exists a rapid change of the first derivative 
around the origin, which is a characteristic of a first-order phase 
transition, the change does not seem to develop by taking $t$ larger.
This suggests that the line $0<x_1<\frac{1}{2}$ 
in \eq{eq:nonperttrans} is merely a crossover line, but not a phase transition line. 
This might be related to the fact that the line vanishes at $x_1=\frac{1}{2}$ in Figure \ref{fig:RGrev}, 
and the two phases on both sides of the line are connected. 
We could not numerically find
any particular features of the grand partition function around $x_2=0$ for $x_1>\frac{1}{2}$.
 
In Figure \ref{fig:RGrev}, the flow diagram seems to show that the line $(x_1,x_2)=(\frac{1}{2}, \hbox{arbitrary})$ should 
also be considered to be a phase transition line. However, as will be explained in 
Section \ref{sec:tensormodel}, the line is a locus of gauge singularities and is not physical.
In fact, we could not see any particular behaviors around the line in the numerical analysis.

\section{The Hamiltonian vector field of the canonical tensor model}
\label{sec:tensormodel}
The main purpose of the present paper is to point out that the phase transition lines of the Ising model 
on random networks of trivalent vertices, which have been numerically studied in the previous sections, 
are in remarkable agreement with what can be derived from the Hamiltonian vector field of 
the canonical tensor model. 
The canonical tensor model has been developed in a series of papers 
\cite{Sasakura:2011sq,Sasakura:2012fb,Sasakura:2013gxg,Sasakura:2013wza}
by one of the current authors as a constructive approach to quantum gravity. 

The dynamical variables of the canonical tensor model are most generally taken to be Hermitian tensors with three
indices. 
In the current application, however, it is enough to assume the tensors with three indices to be symmetric
and real for simplicity, 
since $M$ in \eq{eq:partdef} is so.  
This simpler setting has also been employed in \cite{Sasakura:2013wza}, 
which is followed in the presentation below.

The canonical tensor model has kinematical symmetries which are actually the same as the orthogonal and 
scale transformations, \eq{eq:invortho} and \eq{eq:invscale}, of the statistical system.
The generators of these gauge symmetries are respectively expressed by
\[
&{\cal J}_{[ab]}=\frac{1}{2}\left( P_{acd}M_{bcd}-P_{bcd}M_{acd} \right), \\ 
&{\cal D}=-\frac{1}{3} M_{abc} P_{abc},
\]
where we have introduced the conjugate momentum to $M$ by
\[
&\{ M_{abc},P_{def}\}=\frac{1}{2}\sum_{\sigma} \delta_{a\,\sigma(d)}\delta_{b\,\sigma(e)}\delta_{c\,\sigma(f)},\\
&\{M_{abc},M_{def}\}=\{P_{abc},P_{def} \}=0.
\]
Here $\{\cdot,\cdot\}$ denotes the Poisson bracket, the summation is over all the permutations of $d,e,f$,
and $[\ ]$ symbolically represents the anti-symmetric character,  ${\cal J}_{[ab]}=-{\cal J}_{[ba]}$.
The orthogonal and scaling gauge transformations are 
generated by $\{\cdot ,{\cal J}_{[ab]}\}$ and $\{\cdot ,{\cal D}\}$, respectively.
The canonical tensor model has been constructed so that these symmetry generators and ``local" Hamiltonians form a 
first-class constraint algebra, which defines the canonical tensor model as a totally constrained system
in a consistent manner.
In fact,
as discussed in \cite{Sasakura:2011sq}, 
the algebraic structure of the constraints is very similar to the first-class constraint algebra 
in the ADM formalism of the general relativity 
\cite{Arnowitt:1962hi,Dirac:1958sq,Dirac:1958sc,DeWitt:1967yk,Hojman:1976vp},
and there exists a formal limiting procedure to take the structure of the canonical tensor model to 
that of general relativity.

In \cite{Sasakura:2012fb}, it is shown that the above algebraic condition as well as some other physically 
reasonable assumptions determine uniquely the Hamiltonian constraints as
\[
{\cal H}_a =M_{abc}M_{bde}P_{cde}.
\]
For a given vector ${\cal N}$, these Hamiltonian constraints generate a flow in the space of $M$ as
\[
\delta M_{abc}=\{ M_{abc}, {\cal H}_d\} {\cal N}_d =
{\cal N}_d M_{dae} M_{ebc}+ {\cal N}_d M_{dbe} M_{eca}+ {\cal N}_d M_{dce} M_{eab}.
\label{eq:flow}
\]

In the present application of this paper, we take the following assumptions for the choice of ${\cal N}$: 
${\cal N}$ is a homogeneous function of $M$, and is covariant under the orthogonal transformation \eq{eq:invortho}.
These properties make the flow consistent with the gauge invariance \eq{eq:invortho} and \eq{eq:invscale}.
Then the simplest choice is
\[
{\cal N}_a=M_{abb}.
\label{eq:simpleN}
\]

To compare the flow with the study of the Ising model in the previous sections,
we will take $N=2$ and project the flow \eq{eq:flow} with \eq{eq:simpleN} 
onto the gauge-fixing surface \eq{eq:gauge}. 
On the gauge-fixing surface, the vector fields generated by the gauge transformations and the Hamiltonian
constraints are respectively expressed as 
\[
&{\cal D} \propto 
\frac{\partial}{\partial M_{111}}+x_1 \frac{\partial}{\partial x_1}+x_2 \frac{\partial}{\partial x_2}, \\
&{\cal J}_{[12]}\propto
(1-2x_1)\frac{\partial}{\partial M_{112}}-x_2 \frac{\partial}{\partial x_1}+3 x_1 \frac{\partial}{\partial x_2}, \\
&{\cal H}\propto
\left(
3\frac{\partial}{\partial M_{111}}+ x_1(1+2x_1) \frac{\partial}{\partial x_1}+3 x_1 x_2 \frac{\partial}{\partial x_2}, \right.
\nonumber\\
&\ \ \ \ \ \ \ \ \ \ \ \ \ \ \ \ \ \ \ \ \left. x_1(1+2 x_1)\frac{\partial}{\partial M_{112}}+  
3 x_1 x_2 \frac{\partial}{\partial x_1}+3 ({x_1}^2+ {x_2}^2) \frac{\partial}{\partial x_2}\right), 
\]
where we have ignored some irrelevant overall constant factors. 
The components of $M_{111}$ and $M_{112}$
in the Hamiltonian vector fields are the directions going out of the gauge surface \eq{eq:gauge}. 
These out-going components can be pulled back to the gauge surface 
by the gauge transformations generated by ${\cal D}$ and ${\cal J}_{[12]}$.
This is equivalent to solving ${\cal D}={\cal J}_{[12]}=0$ for 
$\frac{\partial}{\partial M_{111}},\ \frac{\partial}{\partial M_{112}}$ and substituting them into ${\cal H}$.
Then, we obtain the Hamiltonian vector fields projected on the gauge surface as
\[
{\cal H}^{proj}\propto \left(
2 x_1 ( x_1-1)
\frac{\partial}{\partial x_1}+3 x_2 (x_1-1) \frac{\partial}{\partial x_2}, 
\frac{4 x_1 x_2 ( x_1-1)}{2 x_1-1} \frac{\partial}{\partial x_1}+\frac{3 (4 {x_1}^3+2 x_1 {x_2}^2-{x_2}^2)}{2 x_1-1} \frac{\partial}{\partial x_2}\right). 
\label{eq:projH}
\]
Finally, by considering the direction, ${\cal N}=(1+x_1,x_2)$, which is \eq{eq:simpleN} in the gauge \eq{eq:gauge}, 
for \eq{eq:projH}, one obtains 
\[
{\cal N}_a {\cal H}_a^{proj} \propto
2 (x_1-1) x_1 \left(1 + x_1 + \frac{2 x_2^2}{2 x_1-1}\right) \frac{\partial}{\partial x_1}+ 
\frac{3 x_2 (1 - x_1^2 + 6 x_1^3 - x_2^2 + 2 x_1(x_2^2-1))}{2 x_1-1}\frac{\partial}{\partial x_2},
\]
which is the vector field in Figure \ref{fig:rgflow}.

Note that, in the above process, there appears a locus of singularities, $1-2x_1=0$, associated with
solving for $\frac{\partial}{\partial M_{112}}$ from ${\cal J}_{[12]}=0$.
One can check that the locus is gauge dependent, 
and therefore should be unphysical.

It would be interesting to see how general the flow diagram is under the change of ${\cal N}$. 
For instance, under the assumptions mentioned above, we can equally consider
\[
{\cal N}_a=M_{abc}M_{bde}M_{cde}.
\]
However, we have not found any qualitative changes of the flow diagram 
from the case with the simplest choice \eq{eq:simpleN}.
And this was so also for some other choices.

\section{Summary and future prospects}
In this paper, we have shown the close relationship between the canonical tensor model and 
a statistical system on random networks of trivalent vertices. 
This is explicitly studied for the case of $N=2$, which corresponds to the Ising model on random networks.
We have seen that the phase structure of the Ising model agrees with the 
prediction from the canonical tensor model with $N=2$, made by regarding 
the Hamilton vector field of the canonical tensor model as the renormalization group flow of the Ising model.
Along the way, we have discussed a general procedure to obtain the exact free energy of the statistical system
in the thermodynamic limit, and 
have actually obtained an explicit exact expression for the Ising model on random
networks.
We have also provided a non-perturbative definition of the grand partition function which sums up
the partition functions of the statistical system on random networks of any number of vertices.
We have numerically investigated it for $N=2$, and have seen that, 
in a non-perturbative extreme limit, it seems to develop singularities, the locations of which 
agree with the prediction from the canonical tensor model.

An important by-product of this paper is that a mean-field-like method has been 
shown to give the exact free energy of the statistical system on random networks 
in the thermodynamic limit.
This is a rather straightforward consequence from the concise expression of the partition function
in terms of integrals.
One would in principle be able to analyze the phase structure rather easily by the mean-field-like method 
for more complicated cases with $N>2$.
New aspects would be expected to appear.

The grand partition function defined in \eq{eq:partdef} with the integration contour \eq{eq:nonpertcontour}
as well as $Z_n$ in \eq{eq:pert}
would be useful as the basic tools for mathematical investigations of the statistical system on random networks.
They are mathematically rigorously defined, and  
their rather simple expressions would be convenient for obtaining mathematically rigorous arguments
to determine rigorously the thermodynamic properties of the statistical system on random networks.
In particular, the numerical study of the singular properties of the grand partition function in Section \ref{sec:numnon}
must be improved by rigorous analytical arguments. 
It would also be straightforward to generalize the (grand) partition function to the cases of 
networks of vertices with higher degrees, which are commonly discussed in the literature.

An interesting suggestion of this paper is that there might exist a renormalization group-like procedure 
for the statistical system on random networks.  
This sounds rather curious, since the locality of a system is usually a necessary ingredient in the coarse-graining process,
but random networks do not have this property.
On the other hand, the consistency between the Hamiltonian vector field of the canonical tensor model 
and the phase structure seems to suggest the existence of such a process.
It would be highly interesting to find the correct meaning of the flow.

The initial motivation of the present work is to use the statistical model to understand the quantum dynamics 
of the canonical tensor model. 
An important goal is to solve the Wheeler-DeWitt equation of the canonical tensor model.
Though this can be done in a direct manner for the simplest non-trivial case of $N=2$ \cite{Sasakura:2013wza},
an efficient method is necessary for larger $N$.
Since the grand partition function has the same symmetries as the kinematical ones of 
the canonical tensor model, as well as properties reflecting the Hamiltonian vector field, 
this could be used as a kind of basis of solving the Wheeler-DeWitt equation.
Another more physically interesting possibility is that the phase structure of the statistical model may be 
useful in understanding the qualitative behavior of the dynamics of the canonical tensor model.
It is considered that the Universe has experienced some phase transitions after its birth, and even the 
birth itself might have been a phase transition. It would be especially interesting if we can map the phase transitions
of the statistical system to the quantum dynamics of the canonical tensor model, or even the Universe itself.

\section*{Acknowledgment}
We are very grateful to Bernard Raffaelli for useful discussions at the initial stage of this work.


\begin{thebibliography}{99}

%\cite{Garay:1994en}
\bibitem{Garay:1994en} 
  L.~J.~Garay,
  ``Quantum gravity and minimum length,''
  Int.\ J.\ Mod.\ Phys.\ A {\bf 10}, 145 (1995)
  [gr-qc/9403008].
  %%CITATION = GR-QC/9403008;%%
  
%%%%%%%%% Tensor Models %%%%%%%%%%%%%%%%%%%%%%%%%%
\bibitem{ambjorn}
  J.~Ambjorn, B.~Durhuus and T.~Jonsson,
  ``Three-Dimensional Simplicial Quantum Gravity And Generalized Matrix
  Models,''
  Mod.\ Phys.\ Lett.\ A {\bf 6}, 1133 (1991).
  %%CITATION = MPLAE,A6,1133;%%
  
\bibitem{sasakura}
N.~Sasakura,
``Tensor model for gravity and orientability of manifold,"
Mod.\ Phys.\ Lett.\ A {\bf 6} 2613 (1991).
  %%CITATION = MPLAE,A6,2613;%%

\bibitem{godfrey}
  N.~Godfrey and M.~Gross,
  ``Simplicial Quantum Gravity In More Than Two-Dimensions,''
  Phys.\ Rev.\ D {\bf 43}, 1749 (1991).
  %%CITATION = PHRVA,D43,1749;%%

%%%  
  
  
   %\cite{Sasakura:2011sq}
\bibitem{Sasakura:2011sq} 
  N.~Sasakura,
  ``Canonical tensor models with local time,''
  Int.\ J.\ Mod.\ Phys.\ A {\bf 27}, 1250020 (2012)
  [arXiv:1111.2790 [hep-th]].
  %%CITATION = ARXIV:1111.2790;%% 
  
%\cite{Sasakura:2012fb}
\bibitem{Sasakura:2012fb} 
  N.~Sasakura,
  ``Uniqueness of canonical tensor model with local time,''
  Int.\ J.\ Mod.\ Phys.\ A {\bf 27}, 1250096 (2012)
  [arXiv:1203.0421 [hep-th]].
  %%CITATION = ARXIV:1203.0421;%%

%\cite{Sasakura:2013gxg}
\bibitem{Sasakura:2013gxg} 
  N.~Sasakura,
  ``A canonical rank-three tensor model with a scaling constraint,''
  Int.\ J.\ Mod.\ Phys.\ A {\bf 28}, 1 (2013)
  [arXiv:1302.1656 [hep-th]].
  %%CITATION = ARXIV:1302.1656;%%
  
  %\cite{Sasakura:2013wza}
\bibitem{Sasakura:2013wza} 
  N.~Sasakura,
  ``Quantum canonical tensor model and an exact wave function,''
  Int.\ J.\ Mod.\ Phys.\ A {\bf 28}, 1350111 (2013)
  [arXiv:1305.6389 [hep-th]].
  %%CITATION = ARXIV:1305.6389;%%    
  
  %\cite{Sasakura:2011ma}
\bibitem{Sasakura:2011ma} 
  N.~Sasakura,
  ``Tensor models and 3-ary algebras,''
  J.\ Math.\ Phys.\  {\bf 52}, 103510 (2011)
  [arXiv:1104.1463 [hep-th]].
  %%CITATION = ARXIV:1104.1463;%%  
  
   %\cite{Arnowitt:1962hi}
\bibitem{Arnowitt:1962hi} 
  R.~L.~Arnowitt, S.~Deser and C.~W.~Misner,
  ``The Dynamics of general relativity,''  gr-qc/0405109.  
  %%CITATION = GR-QC/0405109;%%

%\cite{Dirac:1958sq}
\bibitem{Dirac:1958sq} 
  P.~A.~M.~Dirac,
  ``Generalized Hamiltonian dynamics,''
  Proc.\ Roy.\ Soc.\ Lond.\ A {\bf 246}, 326 (1958).
  %%CITATION = PRSLA,A246,326;%%

%\cite{Dirac:1958sc}
\bibitem{Dirac:1958sc} 
  P.~A.~M.~Dirac,
  ``The Theory of gravitation in Hamiltonian form,''
  Proc.\ Roy.\ Soc.\ Lond.\ A {\bf 246}, 333 (1958).
  %%CITATION = PRSLA,A246,333;%%
  
%\cite{DeWitt:1967yk}
\bibitem{DeWitt:1967yk} 
  B.~S.~DeWitt,
  ``Quantum Theory of Gravity. 1. The Canonical Theory,''  Phys.\ Rev.\  {\bf 160}, 1113 (1967).  
  %%CITATION = PHRVA,160,1113;%%

%\cite{Hojman:1976vp}
\bibitem{Hojman:1976vp} 
  S.~A.~Hojman, K.~Kuchar and C.~Teitelboim,
  ``Geometrodynamics Regained,''  Ann. Phys.\  {\bf 96}, 88 (1976).  
  %%CITATION = APNYA,96,88;%%   
  
 %\cite{Sasakura:2014gia}
\bibitem{Sasakura:2014gia} 
  N.~Sasakura and Y.~Sato,
  ``Interpreting canonical tensor model in minisuperspace,''
    Phys.\ Lett.\ B {\bf 732}, 32 (2014) [arXiv:1401.2062 [hep-th]]. 

%\cite{Durhuus:2011au}
\bibitem{Durhuus:2011au} 
  B.~Durhuus and G.~M.~Napolitano,
  ``Generic Ising Trees,''
  J.\ Phys.\  {\bf 45}, 185004 (2012)
  [arXiv:1107.2964 [cond-mat.stat-mech]].
  %%CITATION = ARXIV:1107.2964;%%  
  
  
%\cite{Kazakov:1986hu}
\bibitem{Kazakov:1986hu} 
  V.~A.~Kazakov,
  ``Ising model on a dynamical planar random lattice: Exact solution,''
  Phys.\ Lett.\ A {\bf 119}, 140 (1986).
  %%CITATION = PHLTA,A119,140;%%
  
%\cite{Boulatov:1986sb}
\bibitem{Boulatov:1986sb} 
  D.~V.~Boulatov and V.~A.~Kazakov,
  ``The Ising Model on Random Planar Lattice: The Structure of Phase Transition and the Exact Critical Exponents,''
  Phys.\ Lett.\ B {\bf 186}, 379 (1987).
  %%CITATION = PHLTA,B186,379;%%

%\cite{Bonzom:2011ev}
\bibitem{Bonzom:2011ev} 
  V.~Bonzom, R.~Gurau and V.~Rivasseau,
  ``The Ising Model on Random Lattices in Arbitrary Dimensions,''
  Phys.\ Lett.\ B {\bf 711}, 88 (2012)
  [arXiv:1108.6269 [hep-th]].
  %%CITATION = ARXIV:1108.6269;%%
  
 %\cite{Gurau:2009tw}
\bibitem{Gurau:2009tw} 
  R.~Gurau,
  ``Colored Group Field Theory,''
  Commun.\ Math.\ Phys.\  {\bf 304}, 69 (2011)
  [arXiv:0907.2582 [hep-th]].
  %%CITATION = ARXIV:0907.2582;%%
  %95 citations counted in INSPIRE as of 17 Jan 2014 
  
  
 %\cite{Gurau:2011xp}
\bibitem{Gurau:2011xp} 
  R.~Gurau and J.~P.~Ryan,
  ``Colored Tensor Models - a review,''
  SIGMA {\bf 8}, 020 (2012)
  [arXiv:1109.4812 [hep-th]].
  %%CITATION = ARXIV:1109.4812;%% 

%\cite{Bachas:1994qn}
\bibitem{Bachas:1994qn} 
  C.~Bachas, C.~de Calan and P.~M.~S.~Petropoulos,
  ``Quenched random graphs,''
  J.\ Phys.\ A {\bf 27}, 6121 (1994)
  [hep-th/9405068].
  %%CITATION = HEP-TH/9405068;%%
  %14 citations counted in INSPIRE as of 02 Feb 2014
  
  \bibitem{johnston}
D.~A.~Johnston and P.~Plechac,
``Equivalence of ferromagnetic spin models on trees and random graphs,''
J. Phys. A: Math. Gen. {\bf 31} (1998) 475-482.

%\cite{Dorogovtsev:2002ix}
\bibitem{Dorogovtsev:2002ix} 
  S.~N.~Dorogovtsev, A.~V.~Goltsev and J.~F.~F.~Mendes,
  ``Ising model on networks with an arbitrary distribution of connections,''
  Phys.\ Rev.\ E {\bf 66}, 016104 (2002)
  [cond-mat/0203227].
  %%CITATION = COND-MAT/0203227;%%
  
\bibitem{leone}
M.~Leone, A.~V\'azquez, A.~Vespignani and R.~Zecchina,
``Ferromagnetic ordering in graphs with arbitrary degree distribution,"
Eur.\ Phys.\ B {\bf 28}, 191 (2002) 
[cond-mat/0203416].
%%CITATION = COND-MAT/0203416;%%

  
\bibitem{whittle}
P.~Whittle, 
``Fields and flows on random graphs". 
In {\it Disorder in Physical Systems}, eds. G.~R.~Grimmett and D.~Welsh
(Oxford, Oxford University Press, 1990), p.337.

\bibitem{dembo}  
A.~Dembo, A.~Montanari, A.~Sly and N.~Sun, 
``The replica symmetric solution for Potts models on d-regular graphs",
arXiv:1207.5500 [math.PR]. 


%\cite{Sasakura:2014yoa}
\bibitem{Sasakura:2014yoa} 
  N.~Sasakura and Y.~Sato,
  ``Exact free energies of statistical systems on random networks,''
  arXiv:1402.0740 [hep-th].
  %%CITATION = ARXIV:1402.0740;%%
  
  
   \end{thebibliography}
  \end{document}